# Nano-Sim: A Step Wise Equivalent Conductance based Statistical Simulator for Nanotechnology Circuit Design

Bharat Sukhwani, Uday Padmanabhan, Janet M. Wang
Electrical and Computer Engineering, University of Arizona, Tucson.

**Abstract:**
*New nanotechnology based devices are replacing CMOS devices to overcome CMOS technology's scaling limitations. However, many such devices exhibit non-monotonic I-V characteristics and uncertain properties which lead to the negative differential resistance (NDR) problem and the chaotic performance. This paper proposes a new circuit simulation approach that can effectively simulate nanotechnology devices with uncertain input sources and negative differential resistance (NDR) problem. The experimental results show a 20-30 times speedup comparing with existing simulators.*

## 1. Introduction

Due to the increasing circuit complexities and scaling limits of the CMOS devices, new devices, such as resonant tunneling diodes (RTD), resonant tunneling transistors (RTT) and carbon nanotubes (CNT), are being investigated to replace the traditional CMOS devices. Different from the existing CMOS devices, the new devices exhibit non-monotonic I-V characteristics which consist of multiple peaks and valleys. In addition, the new devices may also demonstrate strong sensitiveness towards uncertain environmental changes. The non-monotonic I-V characteristics and the response to uncertain changes are the two main issues in the circuit modeling and simulation for the nanotechnologies.

The traditional deterministic circuit simulators, such as SPICE, estimate the nonlinear device performance by the differential conductance technique together with Newton-Raphson (NR) iterations. When used to simulate the non-monotonic I-V characteristics, differential conductance technique introduces negative differential resistance (NDR) problem which either causes oscillations of Newton-Raphson iterations or results in false convergence during transient simulation. When applied to analyzing systems with uncertain sources, especially time variant uncertain sources, SPICE-like deterministic simulators require several hundreds to over thousands of Monte Carlo simulations at each time point. The high computational complexity at each time step makes the traditional circuit simulators unable to analyze practical circuits.

Recent research work attempts to modify the Newton-Raphson method to force it to converge to meaningful solutions. For example, Bhattacharya and Mazumder [1] proposed current stepping and time-step auto reduction schemes to modify the SPICE simulators. Le et al. in [2] proposed a piece-wise linear approach to replace Newton-Raphson iterations. The paper approximated the nonlinear nanodevice by piece-wise linear conductance. By applying an adaptive time step control mechanism together with the current stepping approach, the method generates accurate results within reasonable run time.

This paper presents two new approaches for nanocircuit modeling and analysis. The first approach models the nanodevice as step-wise equivalent conductance (SWEC) to avoid non-linear devices related Newton-Raphson iterations. In cases of non-monotonic I-V characteristics, the new approach always models the devices as positive conductance. Thus, the SWEC technique completely prevents the occurrence of the NDR problem. The second approach predicts the nanocircuit performance with uncertain inputs by a new stochastic integration technique called Euler-Maruyama method (EM). Equivalent to Euler integration approaches in the deterministic differential equations, the EM method numerically integrates the stochastic differential equations in the time domain to approximate the solutions at each time step.

The rest of this paper adheres to the following format. Section 2 summarizes the anticipated characteristics of the nanoelectronic devices and the existing techniques for the simulation of nanocircuits. Section 3 discusses the features and methodology of SWEC and its application to the resonant tunneling diodes. Section 4 explains the EM based performance prediction. Section 5 presents the results and Section 6 concludes this paper.

## 2. Background

### 2.1 Anticipated characteristics

Strong quantum effects in nanotransistors and nanowires manifest themselves in a level of "potentialities" and a level of "actualizations". "Potentialities" point to a probabilistic approach to the modeling of nanodevices and nanowires. "Actualizations" demand an accurate estimation approach to capture the discrete nature of the devices and wires. The proposed two approaches in this paper target both "potentialities" and "actualizations".

**2.1.1. Actualizations.** In most RTT based nanotransistors, for example, the different discrete energy levels of each material within the transistor terminals act as barriers to current flow. Current flows only when a modulated voltage aligns these energy levels. Electrons may then resonate, "tunnel" across the base, and thus provide current flow from the emitter to the collector. The resulting I-V characteristics exhibits multiple peaks with a staircase contour that leads to the negative differential resistance problems. Figure 1(a) shows the collector current ($I_C$) versus the collector-emitter voltage ($V_{CE}$) for the RTT. A similar curve applies to nanowires. Figure 1(b) illustrates the I-V characteristics of an individual carbon nanotube (CNT). The staircase characteristics of the conductance signal confirms that the carbon nanotubes behave as quantum wires.

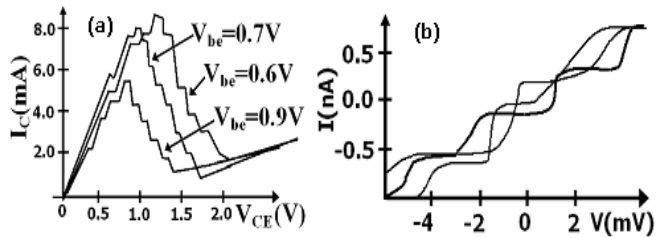

**Figure 1 - I-V curves for (a) RTT (b) CNT [9]**

**2.1.2 Potentialities.** Nanocircuits are sensitive to the environmental changes and have been widely accepted as best candidates for future biomedical sensors. However, the highly sensitive property also leads to uncertain performance. One way to predict the performance is to model the nanocircuit as a system with random inputs.

Conventional circuit simulation tools, such as SPICE, can only handle circuits with deterministic inputs. The current paper is the first one which discusses the possible extension of Euler integration method to Euler-Maruyama integration method.

In the next two sections, we demonstrate the SWEC application in the RTD-based circuit and the EM approach to analyze circuit with random inputs.

## 3. Step Wise Equivalent Conductance Model

### 3.1 NDR problem in Existing Simulators

SPICE-like simulators use Newton-Raphson method to solve nonlinear circuit equations. One important assumption made during these iterations is that the initial guess is close enough to the correct solution of the equation. When the initial guess is far from the correct solution, successive iterations produce oscillatory results even if the circuit is not oscillatory.

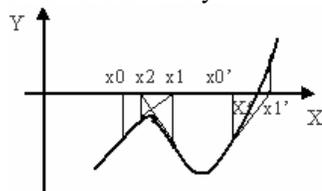

**Figure 2 - Dependence of NR method on initial guess [8]**

Figure 2 shows the dependence of the convergence of Newton-Raphson method on the initial guess. Starting with initial guess x0 leads to oscillations between points x1 and x2 whereas having x0' as the initial guess makes the simulation converge [8].

During transient analysis, SPICE uses the solution at one time point as the initial guess for the next iteration. This strategy works when the applied voltage change is very small. However, if the previous simulation point had a much different applied voltage than the current point, this could also lead to convergence problems. In order to avoid such problems, SPICE employs techniques such as source stepping and device limiting. But, as pointed out by [8], none of these techniques are helpful while simulating circuits having non-monotonic I-V characteristics.

### 3.2 Step Wise Equivalent Conductance Model

Due to the quantum effect, nanotechnology devices exhibit the staircase I-V characteristics shown in Figure 1 (a) and (b). The Step Wise Equivalent Conductance technique approximates the non-linear characteristics of a nanodevice with a stepwise constant conductance that captures the staircase characteristics accurately. It is a non-iterative method and involves computation of the equivalent conductance at each time step. The technique exploits the fact that within a reasonably small time step, every non-linear circuit behaves linearly [4]. It replaces a nonlinear circuit by a linear circuit composed of time-varying conductors. For the integration of the time-varying circuit within one time step, an effective constant conductance is determined for each time-varying conductor. The implicit integration of the equivalent linear time-varying circuit does not require solution of any non-linear equation. The method is consistent, absolutely stable and convergent.

The idea of Step-Wise Equivalent Conductance can be best explained by the nodal equation of a circuit

$$G(t)V(t) + C\dot{V}(t) = bu_s(t) \quad (1)$$

$G$ is the conductance matrix and $C$ is the capacitance matrix. At every time point, the value of devices' equivalent conductance is evaluated and the $G$ matrix is updated. It is important to note that the estimation of the equivalent conductance of the device at every time point is very prone to error and the accuracy of the approach depends on how the time step is chosen. Too large a time step might lead to the failure of implicit integration methods to capture the response of the circuit, while too small a time step hampers the speed of the simulator. To achieve a good tradeoff between the accuracy and speed, we need to adaptively control the time step according to the situation at every time point. The adaptive time step control scheme is explained in detail in the Section 3.4.

For example, the I-V characteristics of a MOS transistor are described as:

$$I_D = \frac{kW}{L}\left((V_{GS} - V_{th})V_{DS} - \frac{V_{DS}^2}{2}\right) \quad \text{if} \quad V_{DS} < V_{GS} - V_{th}$$

$$= \frac{kW}{2L}(V_{GS} - V_{th})^2 \quad \text{if} \quad V_{DS} > V_{GS} - V_{th} \quad (2)$$

where $k$ is the transconductance parameter, $W$ is the effective channel width, and $L$ is the effective channel length of the transistor. The MOS transistor's equivalent conductance ($G(t)$) is the ratio of $I_{DS}$ and $V_{DS}$, evaluated at that time [3].

$$G(t) = \frac{kW}{L}\left((V_{GS}-V_{th}) - \frac{V_{DS}}{2}\right) \quad \text{if} \quad V_{DS} < V_{GS} - V_{th}$$

$$= \frac{kW}{2L}\frac{(V_{GS}-V_{th})^2}{V_{DS}} \quad \text{if} \quad V_{DS} > V_{GS} - V_{th} \quad (3)$$

If the gate voltage ($V_{GS}$) is less than $V_{th}$, the drain current $I_D$ is zero and hence the equivalent conductance $G(t)$ is also zero. The transient analysis evaluates the equivalent conductance of every device at each time point using equation (3). The equivalent conductance remains unchanged until the next point in time, giving rise to a stepwise signal in the time domain.

Because of the possible non-monotone behavior of the device characteristics, both SPICE (based on first order derivative of device I-V characteristics) and ACES (based on a piece wise linear approximation of the device) experience the negative differential resistance problems. SWEC, on the other hand, produces a positive equivalent conductance even for a non-monotone signal. Figure 3 (a) and (b) compare the equivalent conductance definition for piecewise linear (PWL) and stepwise approximations.

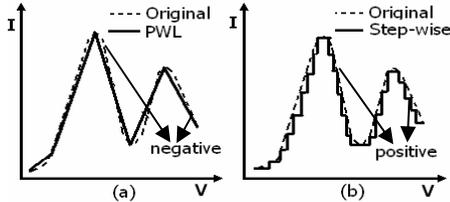

**Figure 3 - Equivalent conductance as per (a) piecewise linear model, and (b) step wise**

To summarize, the SWEC approach only calculates the positive and constant equivalent conductance at each point in time. Therefore it avoids both computationally expensive solution of non-linear equations and the NDR problem.

## 3.3 Application of SWEC to RTDs

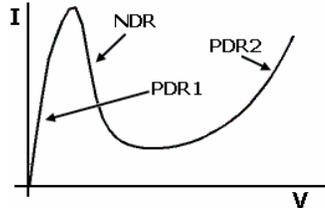

**Figure 4 - RTD I-V Characteristics**

Schulman, De Los Santos, and Chow in [5] provided I-V equation of an RTD:

$$J_1(V) = A\ln\left[\frac{1+e^{(B-C+n_1V)q/kT}}{1+e^{(B-C-n_1V)q/kT}}\right]\left[\frac{\pi}{2} + \tan^{-1}\left(\frac{C-n_1V}{D}\right)\right]$$

$$J_2(V) = H(e^{n_2qV/kT} - 1)$$

$$J(V) = J_1(V) + J_2(V) \quad (4)$$

where $A$, $B$, $C$, $D$, $H$, $n_1$ and $n_2$ are device parameters which define the I-V characteristics of a particular RTD. The I-V characteristics of an RTD are shown in Figure 4. It can be divided into three regions, a first positive differential resistance region (PDR1), a negative differential resistance region (NDR) followed by a second positive differential resistance region (PDR2).

Based on 1$^{st}$ order Taylor expansion, Equation (5) approximates the value of the equivalent conductance of the non-linear circuit elements at the next time point:

$$G_{eq}(n+1) = G_{eq}(n) + \frac{h_n}{2}G'_{eq}(n) \quad (5)$$

where $h_n$ is the value of the time step at $n$ and $G'_{eq}(n)$ is the time derivative of the equivalent conductance evaluated at $n$. The values of $G_{eq}(n)$ and $G'_{eq}(n)$ for an RTD can be evaluated using equation (6) and (7) respectively.

$$G_{eq} = \frac{J_1(V) + J_2(V)}{V}$$

$$= \frac{A\ln\left[\frac{1+\beta}{1+\alpha}\right]\cdot\left[\frac{\pi}{2} + \tan^{-1}\left(\frac{C-n_1V}{D}\right)\right] + H(e^{n_2qV/kT} - 1)}{V}$$

where $\alpha = e^{(B-C-n_1V)\frac{q}{kT}}$ and $\beta = e^{(B-C+n_1V)\frac{q}{kT}}$. (6)

The expression for the equivalent conductance of an RTD (equation (6)) contains the voltage across the RTD as the only time varying quantity. Hence, the time derivative of the equivalent conductance of an RTD can be written as

$$G'_{eq\_RTD} = \frac{dG_{eq\_RTD}}{dt} = \frac{dG_{eq\_RTD}}{dV}\cdot\frac{dV}{dt} \quad (7)$$

where

$$\frac{dG_{eq\_RTD}}{dV} = \frac{1}{V}\left[\frac{n_1qA}{kT}\left(\frac{\beta}{1+\beta} + \frac{\alpha}{1+\alpha}\right)\cdot\left(\frac{\pi}{2} + \tan^{-1}\left(\frac{C-n_1V}{D}\right)\right)\right]$$

$$+ \frac{1}{V}\cdot A\ln\left(\frac{1+\beta}{1+\alpha}\right)\frac{(-Dn_1)}{D^2 + (C-n_1V)^2} + \frac{1}{V}\cdot\frac{qHn_2}{kT}e^{n_2qV/kT} \quad (8)$$

$$- \frac{A\ln\left[\frac{1+\beta}{1+\alpha}\right]\cdot\left[\frac{\pi}{2} + \tan^{-1}\left(\frac{C-n_1V}{D}\right)\right] + H(e^{n_2qV/kT} - 1)}{V^2}$$

$$\frac{dV}{dt} = \frac{V_{RTD}(t_n) - V_{RTD}(t_{n-1})}{h_{n-1}} \quad (9)$$

Figure 5 shows the differential conductance result for an RTD from [1] and stepwise equivalent conductance result from our approach. The differential conductance approach generates negative values of the conductance as the device enters the resistance decreasing region (RDR), whereas the stepwise equivalent conductance approach always generates positive values of the conductance.

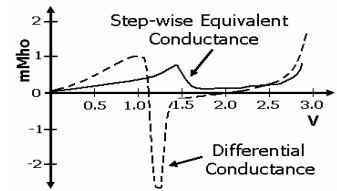

**Figure 5 - RTD conductance as function of applied bias**

## 3.4 Adaptive control of Time Steps

Selection of step size is one of the critical steps of circuit simulation. SWEC employs adaptive control of time step according to the situation at each time point [4]. For a given percent error, we try to maximize the size of the time step so as to speed up the simulations. The percentage local error, $\varepsilon$, at the output at time $t_{n+1}$ is given as

$$\frac{|\Delta V_o(t_n, t_{n+1}) - \Delta V_{o_n}|}{\Delta V_o(t_n, t_{n+1})} \quad (10)$$

where $\Delta V_o(t_n, t_{n+1})$ is the actual change in output voltage and $\Delta V_{o_n}$ is the estimated change by using the equivalent conductance model [4]. For a given value of $\varepsilon$, the constraint on the time step can be calculated.

Figure 6 shows an inverter and its RC equivalent circuit assuming the PMOS to be off. $G\_eq$ is the equivalent conductance of the NMOS. It can be shown that the % local error [4] will be less than $\varepsilon$ if

$$h < \frac{3V_i^0}{\alpha}\varepsilon \quad , \quad h < \frac{C_L}{G^0}\varepsilon \quad (11)$$

where $\alpha = \frac{dV_{in}(t)}{dt} = \frac{dV_{GS}}{dt}$

Experimental results indicate that a good accuracy is achieved if these constraints are satisfied for all the devices in the circuit [4].

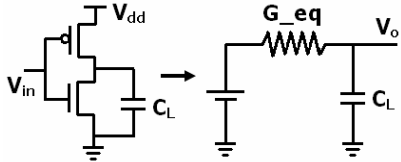

**Figure 6 - Inverter and its RC equivalent circuit**

Thus the time step constraints can be calculated for each transistor and the minimum of those can be used as the next time step, i.e.

$$h_n = \varepsilon * MIN\left(\frac{C_j}{\sum_k G_{jk}(t_n)}, 2\frac{|(V_{GS} - V_t)_i|}{|\alpha_i|}\bigg|_{t_n}\right), \forall i, \forall j \quad (12)$$

$i$ is the index of the transistors which are on and $j$ is the index of the nodes. $C_j$ is the grounded capacitor at node $j$. $\sum_k G_{jk}(t_n)$ is the sum of all the conductance connected to node $j$ at time $t_n$ [4].

## 4. Transient Simulation with Random Inputs

Circuit response with random noise is an active research topic in the analog circuit design area. Because of the uncertain behavior of nanodevices, a few recent papers [11] [12] also discussed the power grid analysis with random current draws from nanodevices. In both research topics, the stochastic differential equation (SDE) model plays an important role in finding the expected value of the circuit response or performance.

However, as pointed out by some recent publications [15], the expected value of the performance can only provide an average performance, while the transient value is also important to most applications. For example, in the power grid analysis case, even though the average voltage drop is zero, if the transient voltage drop at a certain time point exceeds certain constraints, the whole design is still going to fail. Likewise, the transient response of nanocircuit offers real time performance estimation and catches the possible signal integrity problems in the circuits.

Existing simulators only analyze circuits with deterministic inputs, the current paper is the first one discussing the transient simulation technique for nanocircuit with uncertain inputs.

### 4.1 Model the Input as Wiener Process

The state equation for a nanocircuit can be written as

$$G(t)x + C\frac{dx}{dt} = Bu(t) \text{ or } dx = C^{-1}G(t)x dt + C^{-1}Bu(t)dt \quad (13)$$

with $x(0) = x_0$ as the initial value. Here $x$ is the response of the circuit, $G$ and $C$ are the parasitic matrices. Since $G$ is time variant, Equation (13) also includes cases with the nonlinear nanodevices. Assume $u(t)$ represents the random time domain input. Because of its high randomness, $u(t)$ is generally modeled as white noise. Let $dW(t) = u(t)dt$. $W(t)$ is called Wiener process or Brownian motion.

A standard Wiener process over $[0,T]$ is a random variable that depends continuously on $t \in [0,T]$ and satisfies three conditions: 1) $W(0) = 0$ (with probability 1); 2) For $0 \le s < t \le T$, the random variable given by the increment $W(t) - W(s)$ is normally distributed with mean zero and variance $t - s$, i.e. $W(t) - W(s) \sim \sqrt{t-s}N(0,1)$ where $N(0,1)$ denotes a normally distributed random variable with zero mean and unit variance; 3) For $0 \le s < t < u < v \le T$ the increments $W(t) - W(s)$ and $W(v) - W(u)$ are independent. For computational purposes it is useful to consider discretized cases where $W(t)$ is specified at discrete $t$ values. We thus set $dt = T/N$ for some positive integer $N$ and let $W_j$ denote $W(t_j)$ with $t_j = jdt$.

### 4.2 The Euler–Maruyama Method

The Stochastic Differential Equation in Equation (13) can be written in integral form as

$$X(t) = X_0 + \int_0^t C^{-1}G(t)X(t)ds + \int_0^t C^{-1}BdW(s) \quad (14)$$

The solution $X(t)$ is a random variable for each $t$. The second integral on the right-hand side of (14) is to be taken as the Ito integral. The Ito integral defines the stochastic integration operation. The stochastic integration can not be understood as a deterministic ordinary integral [13] [14]. For example, the following two ways of

$\int_0^T h(t)dt$ integration converge to the same solution in the deterministic cases:

$$\sum_{j=0}^{N-1} h(t_j)(t_{j+1} - t_j) \quad \text{and} \quad \sum_{j=0}^{N-1} h(\frac{t_j + t_{j+1}}{2})(t_{j+1} - t_j)$$

By analogy with the above two ways, the stochastic integral $\int_0^T h(t)dW(t)$ can be estimated as in Equation (15) or (16):

$$\sum_{j=0}^{N-1} h(t_j)(W(t_{j+1}) - W(t_j)) \quad (15)$$

$$\sum_{j=0}^{N-1} h(\frac{t_j + t_{j+1}}{2})(W(t_{j+1}) - W(t_j)) \quad (16)$$

Equation (15) and (16) give markedly different answers. Even with $\Delta t \to 0$, the mismatch of the two equations does not go away. This emphasizes the significant difference between deterministic and stochastic integration: that is, we have to be precise about the way the sum is formed. Though the expected value of the results from Equation (15) and (16) are the same, for the purpose of transient performance prediction, each equation leads to different prediction mechanism [14]. In the current paper, we use Equation (15) for stochastic integration. Equation (15) is also referred as Ito integral.

Euler-Maruyama defines a numerical method for solving Equation (13) with Ito integral. The solution $X(t)$ as the random variable arises when we take the zero stepsize limit in the numerical method. It is usual to rewrite (14) in differential equation form as

$$dX(t) = C^{-1}G(t)dt + C^{-1}BdW(t) \quad (17)$$

with $X(0) = X_0$ and $0 \le t \le T$. To apply a numerical method to (17) over [0, T], we first discretize the interval. Let $\Delta t = T/L$ for some positive integer $L$, and $\tau_j = j\Delta t$. Our numerical approximation to $X(\tau_j)$ will be denoted $X_j$. The Euler–Maruyama (EM) method takes the form

$$X_j = X_{j-1} + C^{-1}G(t_{j-1})\Delta t + C^{-1}B(W(t_j) - W(t_{j-1})) \quad (18)$$

where $j = 1, 2, \cdots, L$. To understand where Equation (18) comes from, notice from the integral form Equation (15) that

$$X(t_j) = X(t_{j-1}) + \int_{t_{j-1}}^{t_j} C^{-1}G(s)ds + \int_{t_{j-1}}^{t_j} C^{-1}BdW(s) \quad (19)$$

Each of the three terms on the right-hand side of Equation (18) approximates the corresponding term on the right-hand side of Equation (19). We also note that in the deterministic case ($B \equiv 0$ and $X_0$ constant), Equation (19) reduces to Euler's method. Following the Black-Scholes approach [13] [14], we can predict the peak performance within certain time window. A close analogy to this problem is the stock price prediction.

## 5. Results

### 5.1 DC Analysis:

Figure 7(a) shows the I-V characteristics of the RTD as captured by our approach. The circuit consisted of a series combination of a resistor and an RTD across a voltage source (a voltage divider circuit). The figure also shows the I-V characteristics as captured by our implementation of the Modified Limiting Algorithm (MLA) presented by Bhattacharya and Mazumder [1]. As we can see, our approach is able to capture the negative resistance region of the I-V curve very closely and accurately. Similar curve for a nanowire is shown in Figure 7(b). A range of voltages were applied to the series combination of a nanowire and a resistor and the nanowire current-voltage obtained using SWEC was plotted. This figure conforms well to the I-V characteristics of a carbon nanotube, indicating that SWEC is able to simulate the circuits involving nanowires.

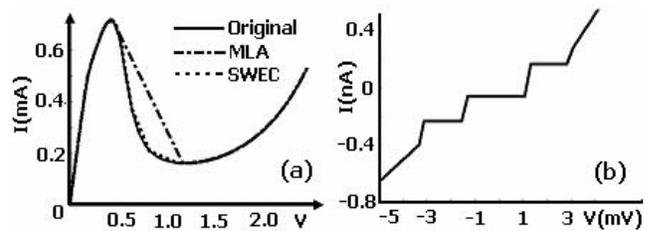

**Figure 7 - I-V characteristics using SWEC (a) RTD (b) Nanowire**

Table I compares the number of floating point operations needed to perform different types of simulations by SWEC and MLA. Due to the unavailability of the MLA code, we present the comparison between SWEC and the implementation of the MLA done by us. As mentioned earlier, SWEC is a non iterative method and thus yields high simulation speed. This is clearly depicted by the results presented in table I.

| Circuit | Type of Simulation | # of Floating Point Operations SWEC | # of Floating Point Operations Impl. of MLA | Speedup of SWEC |
|---|---|---|---|---|
| Series RTD and Resistor | Operating Point | 662 | 1605 | 2.4 |
| RTD Inverter | Operating Point | 1332 (High i/p) 223 (Low i/p) | 2647 (High i/p) 541 (Low i/p) | 2 2.4 |
| RTD D-f/f | Operating Point | 1127 | 1676 | 1.48 |
| Inverter Chain | Operating Point | 5238 (High i/p) 1931 (Low i/p) | 10417 (High i/p) 11443 (Low i/p) | 1.98 5.9 |
| Series RTD and Resistor | DC Sweep (60 points) | 2062 | 61216 | 29.7 |
| Series RTD and Resistor | DC Sweep (500 points) | 17462 | 543352 | 31 |

**Table I - Comparison of DC simulations performance**

### 5.2 Transient Analysis:

We simulated an FET-RTD inverter circuit using the stepwise equivalent conductance technique. The input voltage switches between 0 and 5V. Figure 8(a) shows the circuit and the output obtained at the junction of two RTDs is shown in Figure 8(b). The values of model parameters used for this simulation are as follows: A = 1e-4; B = 2; C =

1.5; D = 0.3; n1 = 0.35; n2 = 0.0172;  H = 1.43e-8. Shown also are the responses obtained by other circuit simulators. As can be seen from Figure 8(c), SPICE3 fails to converge to the correct solution [2]. SWEC generates more accurate response without needing to solve set of non linear equations, thus yielding better results at less computational expense.

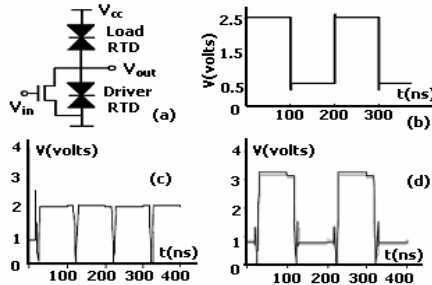

**Figure 8 - (a) FET-RTD Inverter, Output generated by (b) SWEC, (c) SPICE3, (d) ACESn [2]**

Figure 9(a) shows the circuit of an RTD-D flip-flop. Refer [6] for details on the working of the circuit. The circuit was simulated with the clock waveform as shown in Figure 9(b). The data applied and the output obtained is as shown in Figure 9(c). The input waveform switches at t = 300ns and the output waveform switches at the rising edge of clock at t = 350ns. This shows that we could capture the right behavior of the circuit using our stepwise equivalent conductance technique.

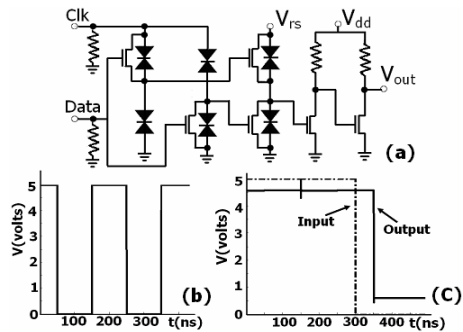

**Figure 9 - (a) RTD-D Flip Flop, (b) Clock Signal, (c) Input and output waveforms**

### 5.3 Performance Prediction:
Figure 10 demonstrates the results from both true solution and EM integration method. The circuit is a time-variant nanoscale transistor with some parasitic RCs. From 0-1ns, we observe a possible performance peak about 0.6 V.

## 6. Conclusions

This paper proposes a new stepwise conductance based statistical simulator. It not only prevents the NDR problems, but also predicts the performance within certain time windows. The proposed simulator has over 20-30 times speedup over the SPICE-like simulator.

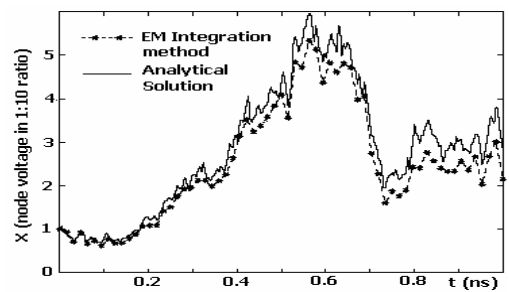

**Figure 10 – Results from EM method and Analytical solution. X is node voltage in 1:10 ratio**